\documentclass[12pt]{article}

\usepackage{latexsym}

\usepackage{graphicx}
\usepackage{dcolumn}

\usepackage{amsmath}

\font\bbbold=msbm10 at 12pt

\newcommand{\sfrac}[2]{ \mbox{$\frac{#1}{#2}$} }

\newcommand{\bbN}{\mbox{\bbbold N}}

\newcommand{\bbZ}{\mbox{\bbbold Z}}

\newcommand{\bq}{\begin{equation}}
\newcommand{\eq}{\end{equation}}

\newcommand{\bqr}{\begin{eqnarray}}
\newcommand{\eqr}{\end{eqnarray}}

\newcommand{\bqrx}{\begin{eqnarray*}}
\newcommand{\eqrx}{\end{eqnarray*}}

\newcommand{\br}{\begin{array}}
\newcommand{\er}{\end{array}}

\newcommand{\blsk}{\baselineskip}

\begin{document}

\pagestyle{plain}
\pagenumbering{arabic}

\setlength{\parindent}{0in}
\setlength{\parskip}{3ex}
\setlength{\footskip}{.6in}


\setlength{\blsk}{.25in}%
\vspace*{.6in}
\begin{center}
On the behavior of quantum walks confined to a cycle coupled with a half
line.
\end{center}
\begin{center}
Forrest Ingram-Johnson,\, Chaobin Liu\footnote{cliu@bowiestate.edu},\, Nelson Petulante \\Department of Mathematics, Bowie State University, \, Bowie, MD 20715 USA\\ 

\end{center}

\begin{abstract}
{When confined to a topological environment consisting of a cycle coupled with a half-line, quantum walks exhibit long-term statistical tendencies which differ dramatically from the tendencies of classical random walks in the same environment.  In particular, as suggested by numerical simulations, the probability distribution of the walker's position resolves, in part, into a non-vanishing distribution on the cycle and, in part, into a ballistic distribution on the half-line. By contrast, for a classical random walk, the probability distribution of the walker's position tends always to vanish on the cycle and to migrate completely to the half-line as a purely diffusive process.}
\end{abstract}%

\section{Introduction}

In recent times, the study of quantum walks has generated a vast volume of publications. For a lively and informative elaboration of the history of quantum walks and their connection to quantum computing and physics, the reader is referred to \cite{K03, K06, K08, VA08, AC09, MB11, VA12} and the references therein. In general, a quantum walk evolves, subject to certain rules of transition, over a topological lattice consisting of a network of nodes, the cardinality of which may be either finite or infinite. For the aims of this report, the following two references are of special interest: \cite{AAKV01}, wherein the topological lattice is a finite cycle and \cite{ABNVW01, K02}, wherein the topological lattice is an infinite line. In \cite{AAKV01}, the long-term behavior of the walker's itinerary fails to converge to any stationary probability distribution. Meanwhile, in \cite{ABNVW01, K02}, the long-term behavior of the walker's itinerary evolves into a pattern resembling an inverted bell asymptotic distribution (formally called ``ballistic propagation"). Some more recent developments on these themes can be found in  \cite{KLS12, LP13, M2013} and the references therein.

In this report, we examine the statistical behavior of a quantum walk confined to a topological lattice consisting of a finite cycle coupled with an infinite half-line. Based on extensive numerical simulations, we observe that the position probability distribution of the walker tends to dichotomize, in part, into a non-stationary distribution on the cycle with non-vanishing terminal probability density and, in part, into a ballistic distribution on the half-line. These findings seem to defy intuitive expectations and appear to deviate dramatically from the known behavior of classical random walks on the same topological setting. In this paper, we do not attempt to formulate theoretical justifications for these phenomena.   

This report is organized as follows. Section 2 is devoted to a formulation of the basic elements. In Section 3, we display the results of numerical simulations of the quantum walks defined in Section 2. In Section 4, we offer some brief conclusions and speculative remarks.

\section{Quantum walks confined to a cycle coupled with a half-line}

Throughout this paper, the topological setting for all walks, both classical and quantum, is a special graph $G$, defined as follows. Viewed as a graphical entity, let $\bbZ_{n}=\{[0],[1],...,[n-1]\}$ denote a cycle with $n$ nodes. Similarly, let $\bbN_{0}$ denote the half-line with the set of non-negative integers as its nodes. Then $G=\bbZ_{n}\cup\bbN_{0}$, where the special node $[0]$ on the $n$-cycle is identified with node $0$ on the half-line. 

Throughout this paper, the transition rules for the quantum ``coin operator" $U$ are defined as follows:

For any non-zero position $x\ge 1$ on the half-line:
$$U |x\downarrow\rangle=\sfrac{\sqrt{2}}{2}|x-1 \downarrow \rangle +\sfrac{\sqrt{2}}{2}|x+1 \uparrow\rangle$$
$$U |x\uparrow\rangle=\sfrac{\sqrt{2}}{2}|x-1 \downarrow\rangle -\sfrac{\sqrt{2}}{2}|x+1 \uparrow\rangle$$

For any position $[x]\ne [0]$ on the cycle:
$$U |[x]L\rangle=\sfrac{\sqrt{2}}{2}|[x-1] L\rangle +\sfrac{\sqrt{2}}{2}|[x+1] R\rangle$$
$$U |[x]R\rangle=\sfrac{\sqrt{2}}{2}|[x-1] L\rangle -\sfrac{\sqrt{2}}{2}|[x+1] R\rangle$$

For the common node $0=[0]$ on the half-line and $n$-cycle:
$$U |0L\rangle=-\sfrac{1}{3}|[n-1] L\rangle +\sfrac{2}{3}|[1] R\rangle+\sfrac{2}{3}|1\uparrow\rangle$$
$$U |0R\rangle=\sfrac{2}{3}|[n-1] L\rangle -\sfrac{1}{3}|[1] R\rangle+\sfrac{2}{3}|1\uparrow\rangle$$
$$U |0\downarrow\rangle=\sfrac{2}{3}|[n-1] L\rangle +\sfrac{2}{3}|[1] R\rangle-\sfrac{1}{3}|1\uparrow\rangle$$

\vskip 0.2in

Suppose the quantum walker is launched with initial state $|\psi_0\rangle$. At any subsequent time $t$, its quantum state is given by $|\psi_t\rangle=U^{t}|\psi_0\rangle=\sum_{x}|\psi_t(x)\rangle$. Let $X_t$ denote the position of the walker at time $t$.  Then, according to standard quantum mechanical conventions, the probability distribution of $X_t$ is identified with $\langle \psi_t(x)|\psi_t(x)\rangle$. 

\vskip0.2in

In what follows, we display the results of numerical simulations of the quantum walk as defined above. For comparison, we  display also the results of numerical simulations of the classical counterpart of the quantum walk defined above. Let $P$ denote the governing probability operator for the classical random walk. The transition rules of $P$ are as follows:

For each $x\ge 1$ on the half-line,
$$P |x\rangle=\sfrac{1}{2}|x-1\rangle +\sfrac{1}{2}|x+1\rangle$$
For each $[x]\ne 0$ on the cycle,
$$P|[x]\rangle=\sfrac{1}{2}|[x-1]\rangle +\sfrac{1}{2}|[x+1]\rangle$$
For $x=0$, 
$$P |0\rangle=\sfrac{1}{3}|[n-1]\rangle +\sfrac{1}{3}|[1]\rangle+\sfrac{1}{3}|1\rangle$$

\section{Numerical simulations}

For all walks, both quantum and classical, whose simulations are displayed below, the topological environment is the graph $G=\bbZ_{25}\cup\bbN_{0}$ consisting of a 25-node cycle coupled with an infinite half-line. A total of eight (8) simulations are displayed: four (4) launched from initial position $[12]$ on the cycle and four (4) launched from initial position $[0]$. The presentation of results is organized as follows:

{\bf Figure 1}: Asymptotic trend on the {\bf half-line} of {\bf quantum walk} launched from {\boldmath $|[12]R\rangle$}\\
{\bf Figure 2}: Asymptotic trend on the {\bf half-line} of {\bf classical walk} launched from {\boldmath $|[12]\rangle$}\\
{\bf Figure 3}: Asymptotic trend on the {\bf cycle} of {\bf quantum walk} launched from {\boldmath $|[12]R\rangle$}\\
{\bf Figure 4}: Asymptotic trend on the {\bf cycle} of {\bf classical walk} launched from {\boldmath $|[12]\rangle$}\\
{\bf Figure 5}: Asymptotic trend on the {\bf half-line} of {\bf quantum walk} launched from {\boldmath $|[0]R\rangle$}\\
{\bf Figure 6}: Asymptotic trend on the {\bf half-line} of {\bf classical walk} launched from {\boldmath $|[0]\rangle$}\\
{\bf Figure 7}: Asymptotic trend on the {\bf cycle} of {\bf quantum walk} launched from {\boldmath $|[0]R\rangle$}\\
{\bf Figure 8}: Asymptotic trend on the {\bf cycle} of {\bf classical walk} launched from {\boldmath $|[0]\rangle$}

\hspace{50pt}
\begin{figure}[h!]
\includegraphics[height=3.0in]{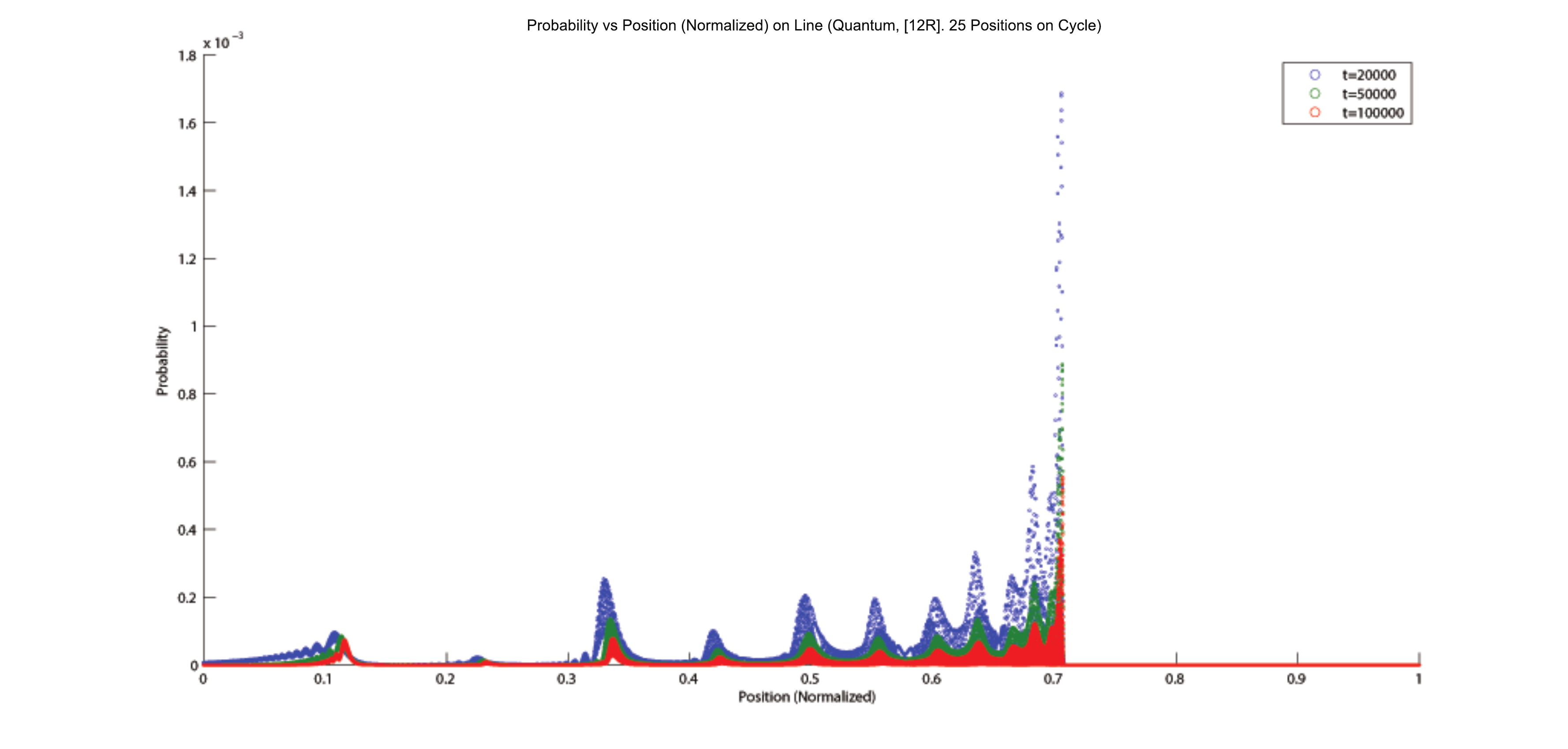}
\caption{\label{fig:wide}When the quantum walker is launched from $|[12]R\rangle$, the probability distribution on the half-line appears to display ballistic behavior.}
\end{figure}

\mbox{}\hspace{50pt}

\begin{figure}[h!]
\includegraphics[height=3.0in]{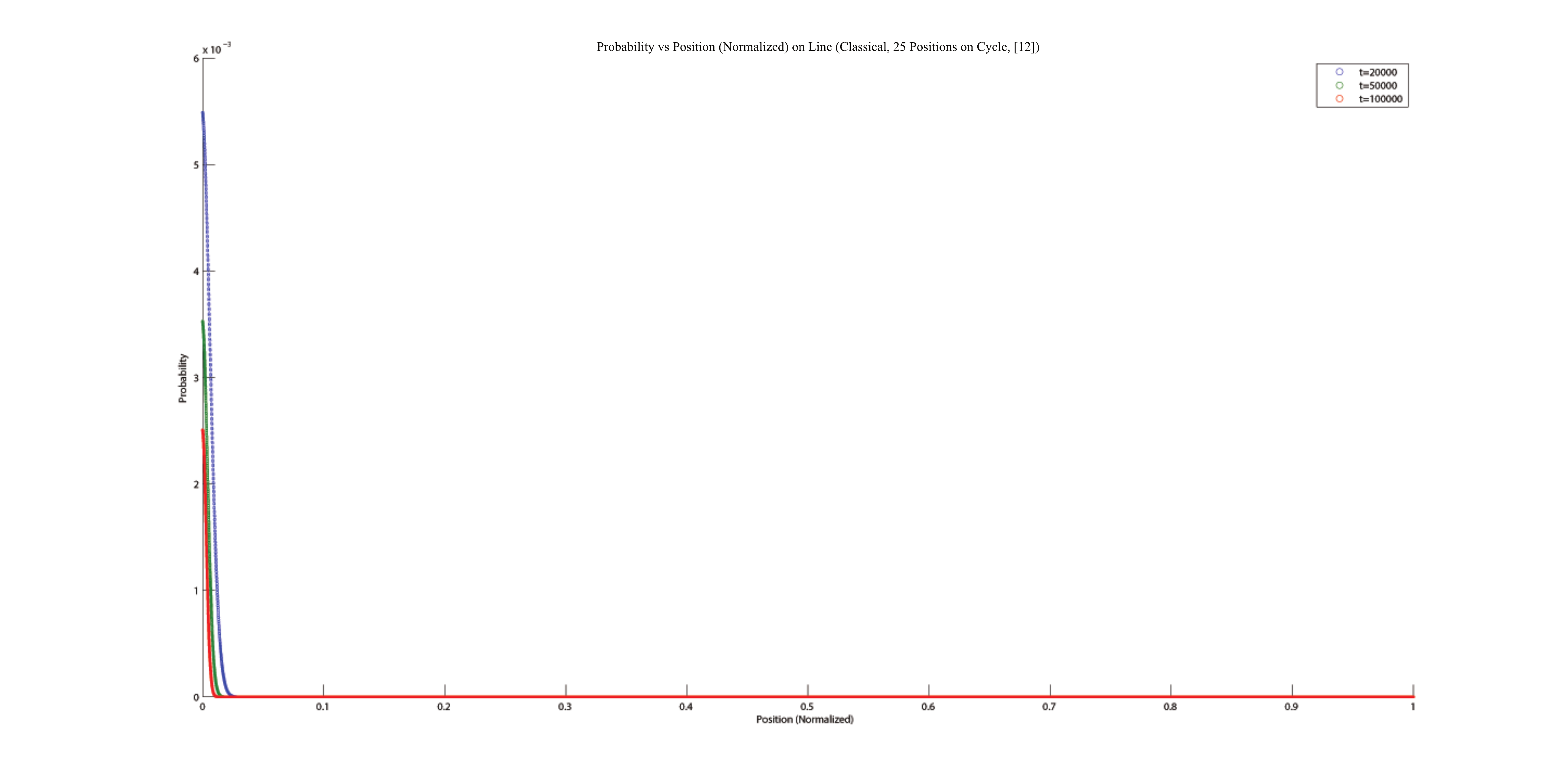}
\caption{\label{fig:wide}When the classical random walker is launched from $|[12]\rangle$, the probability distribution on the half-line exhibits diffusive behavior.}
\end{figure}

\newpage

\begin{figure}[h!]
\includegraphics[height=3.0in]{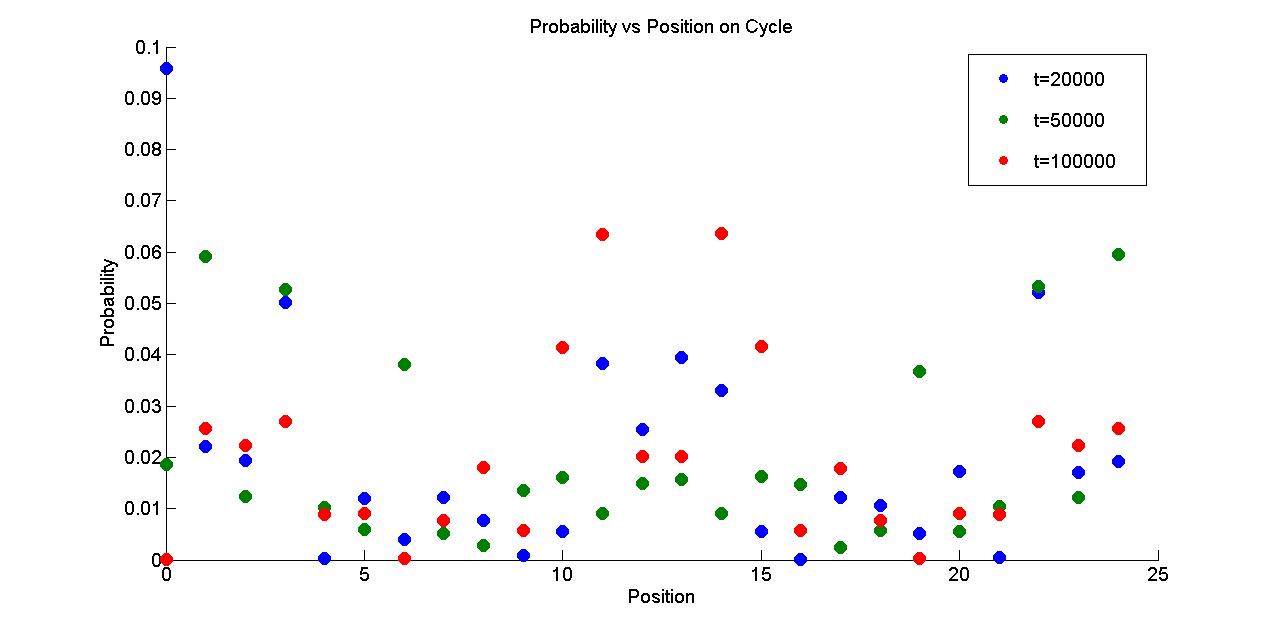}
\caption{\label{fig:wide}When the quantum walker is launched from $|[12]R\rangle$, the probability distribution on the cycle appears quasi-periodic and the total probability on the cycle appears to converge to a non-zero value. The numerical values are illustrated in Table 1 below (see Section 4).}
\end{figure}

\mbox{}\hspace{50pt}

\begin{figure}[h!]
\includegraphics[height=3.0in]{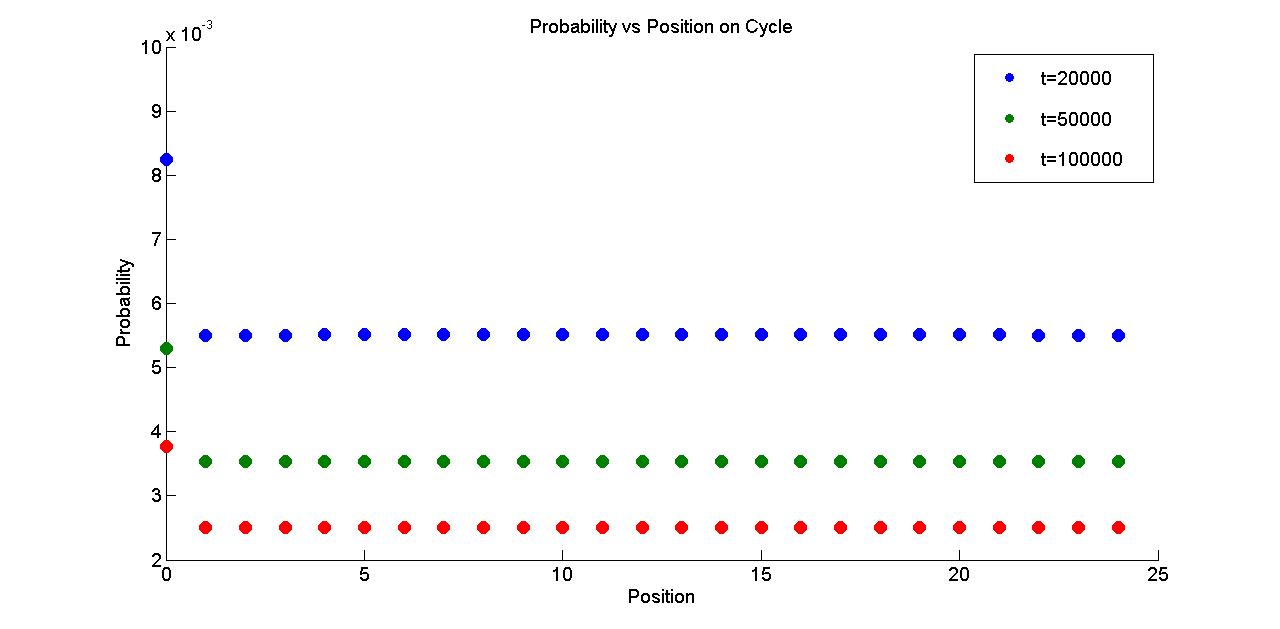}
\caption{\label{fig:wide}When the classical random walker is launched from $|[12]\rangle$, the probability spread on the cycle tends to a uniform distribution whose total probability vanishes as $t$ tends to infinity.  The numerical values are illustrated in Table 2 below (see Section 4).} 
\end{figure}
 
\newpage

\begin{figure}[h!]
\includegraphics[height=3.0in]{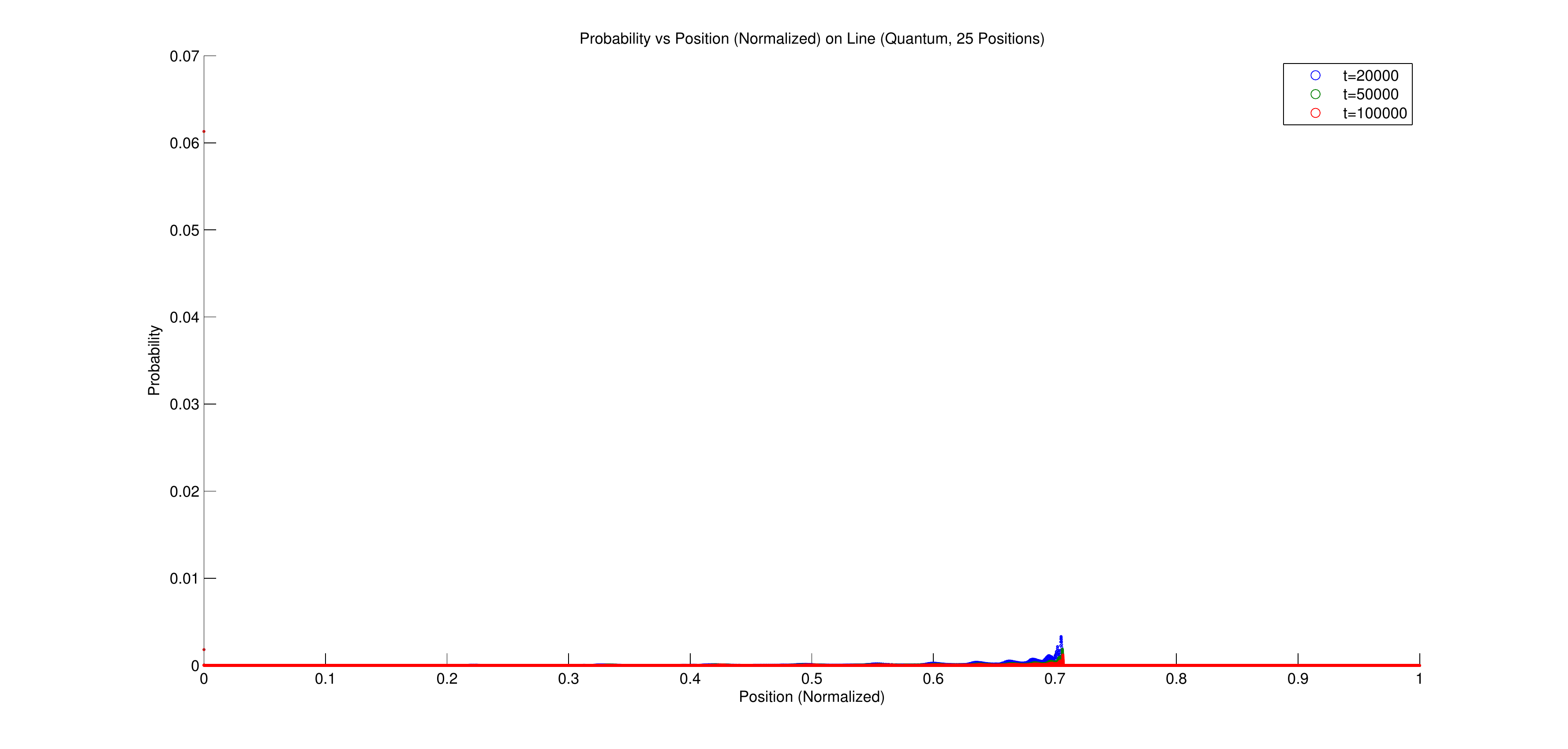}
\caption{\label{fig:wide}When the quantum walker is launched from $|[0]R\rangle$, a spike is evident at the initial position $[0]$ and the probability distribution on the half-line exhibits ballistic behavior.}
\end{figure}

\mbox{}\hspace{50pt}

\begin{figure}[h!]
\includegraphics[height=3.0in]{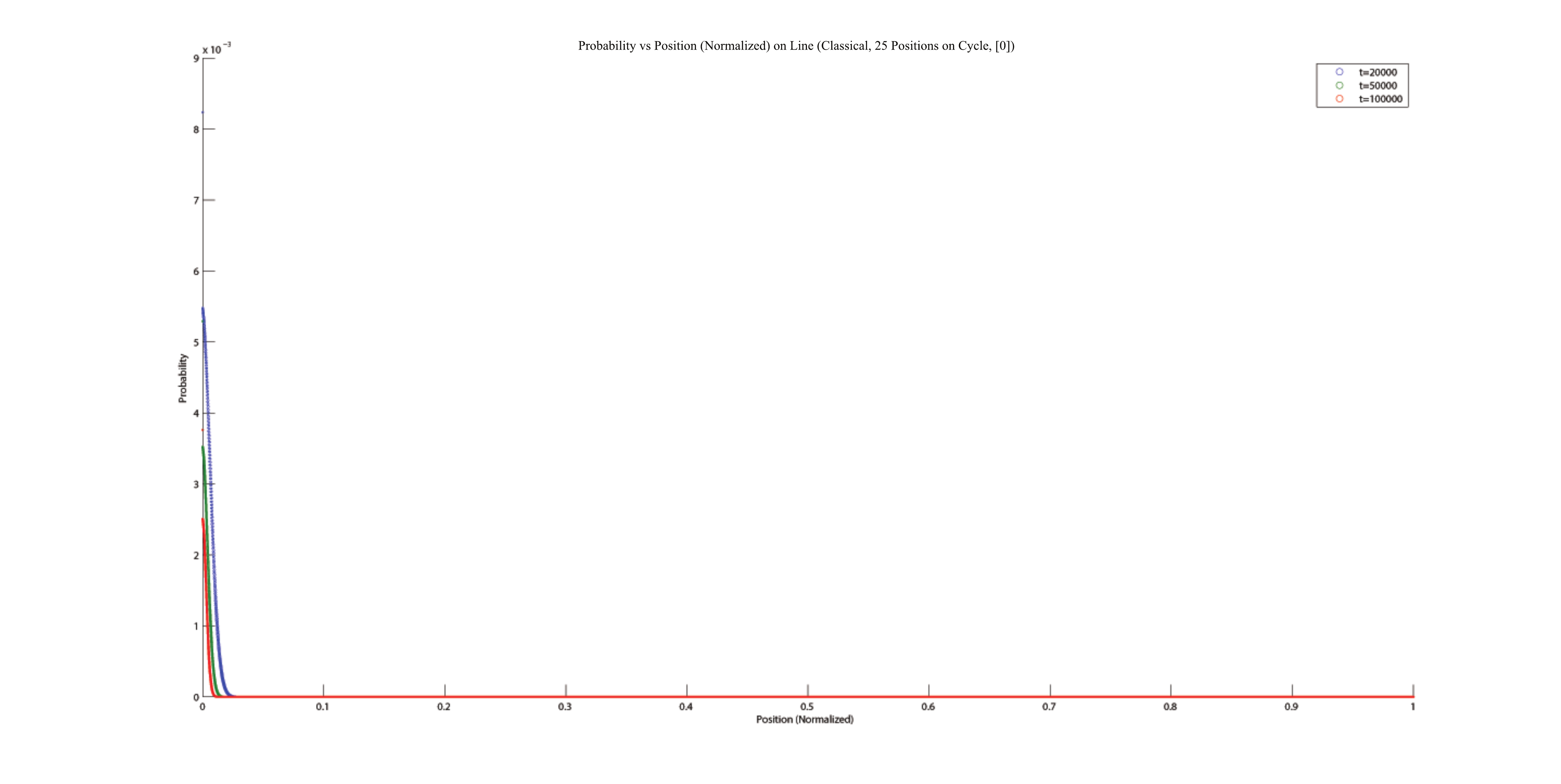}
\caption{\label{fig:wide}When the classical walker is launched from $|[0]\rangle$, the probability distribution on the half-line exhibits diffusive behavior.}
\end{figure}

\newpage

\begin{figure}[h!]
\includegraphics[height=3.0in]{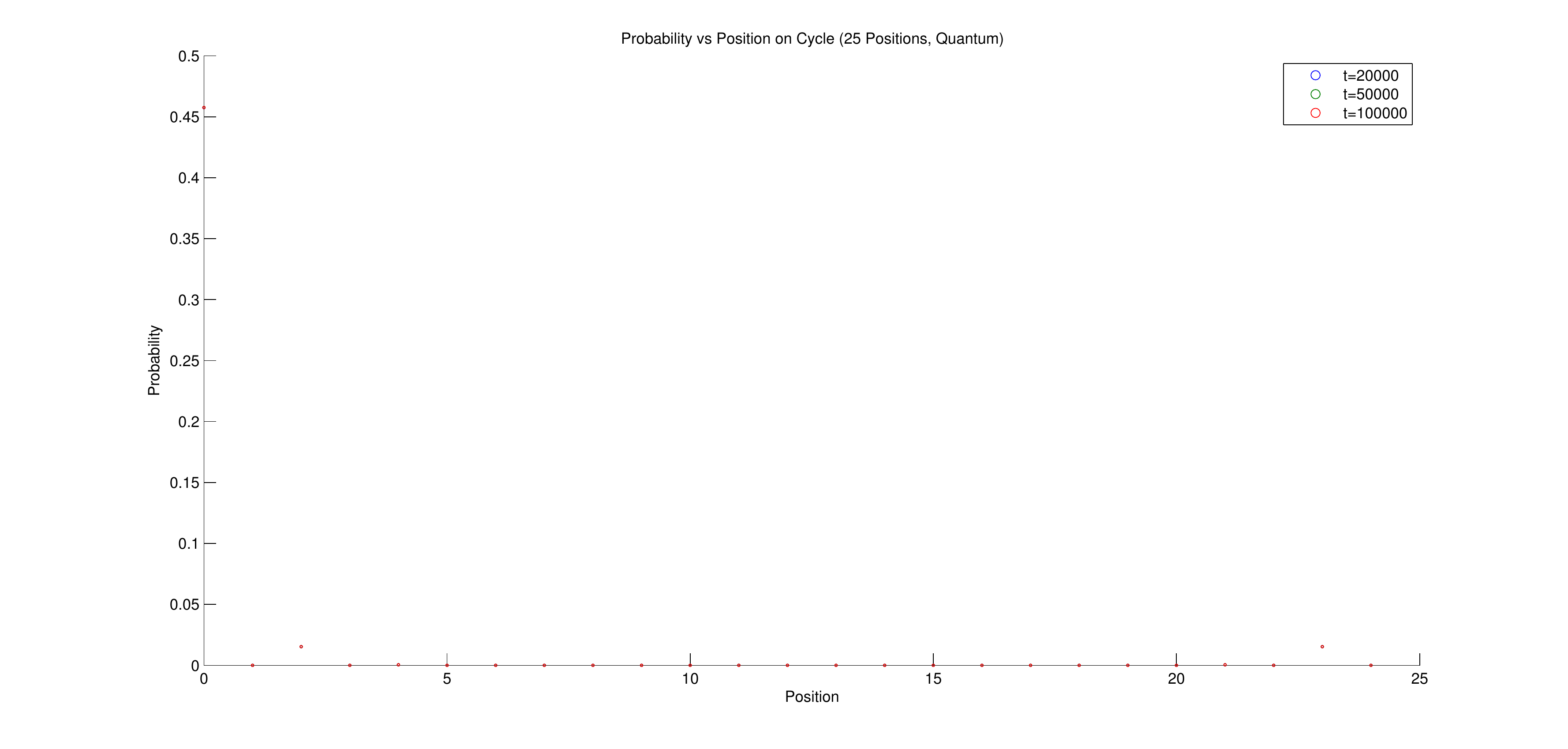}
\caption{\label{fig:wide}When the quantum walker is launched from $|[0]R\rangle$, the probability distribution on the cycle appears quasi-periodic and the total probability on the cycle appears to converge to a positive value. For instance, at times $t=20000, 50000, 100000$,  strong localization at $|0\rangle$ is evident with probability value $\approx .457$.}
\end{figure}

\mbox{}\hspace{50pt}

\begin{figure}[h!]
\includegraphics[height=3.0in]{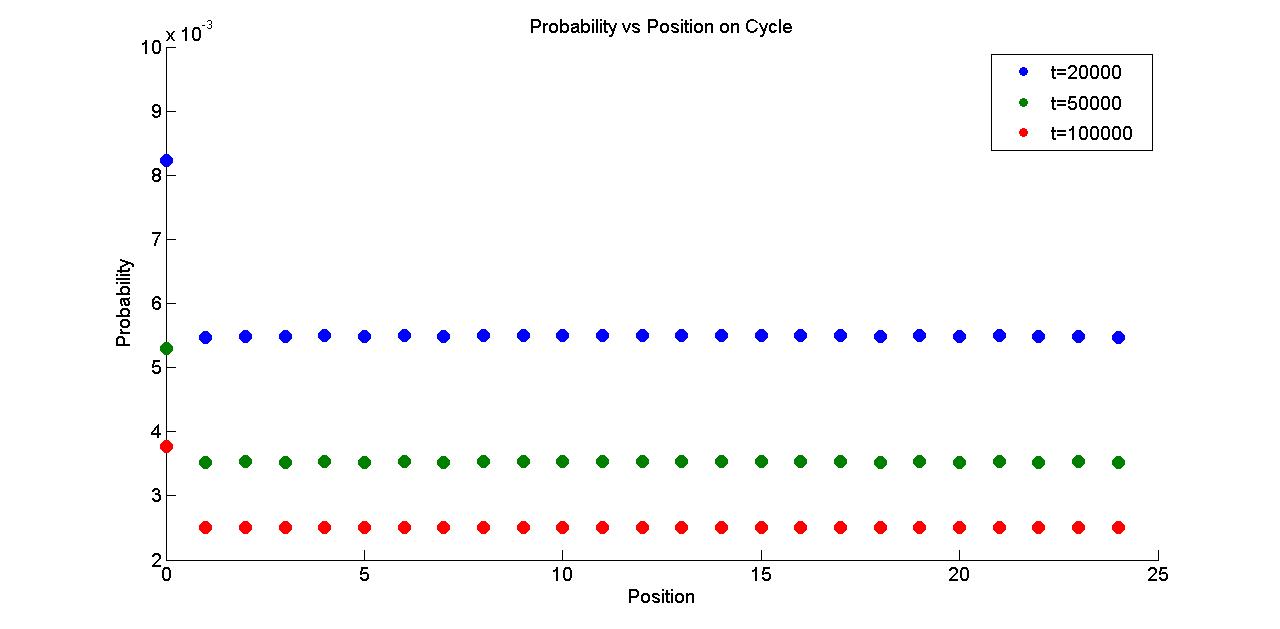}
\caption{\label{fig:wide}When the classical walker is launched from $|[0]\rangle$, the probability distribution on the cycle appears uniform and the total probability on the cycle tends to zero. For instance, at times $t=20000, 50000, 100000$, the corresponding total probability values on the cycle are $.14003, .08998, .06398$ respectively.}
\end{figure}

\newpage

\section{Conclusion and Remarks}
 
The following two tables summarize the data displayed graphically in Figures 3 and 4 above.

\begin{table}[h!]
\caption{Probability Distribution on the cycle for the quantum walk starting at $|[12]R\rangle$} 
\vspace*{20pt}
\centering  
\begin{tabular}{c c c c c} 
\hline\hline                        
t &20000 &50000 & 100000 \\ [0.5ex] 
\hline                  
$\mathrm{Spike\ position}$ &  [0]&  [24]& [14]\\
\hline                  
$\mathrm{Spike\ height}$ &  .09573 & .05955& .06355.\\ 
\hline
$P_{total}$ &.50587&.50012&.50000\\
\hline 
\end{tabular}
\label{table:nonlin} 
\end{table}

\begin{table}[h!]
\caption{Probability Distribution on the cycle for the classical walk starting at $|[12]\rangle$} 
\vspace*{20pt}
\centering  
\begin{tabular}{c c c c c} 
\hline\hline                        
t &20000 &50000 & 100000 \\ [0.5ex] 
\hline                  
$\mathrm{Spike\ height\ at\ [0]}$ & .00825 &.00530  &.00376\\ 
\hline
$P_{total}$ &.14055&.09011&.06403\\
\hline 
\end{tabular}
\label{table:nonlin} 
\end{table}

As suggested by the simulations displayed above, when confined to the special graph $G=\bbZ_{25}\cup\bbN_{0}$, a classical random walk behaves as expected, both intuitively and analytically. The classical walker tends to drift ad infinitum along the half-line with vanishing probability density at every point on the entire graph. By contrast, on the same graph, the quantum walker tends to persist on the cycle ad infinitum with non-vanishing total probability density. No doubt this strange behavior reflects the so-called ``localization" property of quantum phenomena \cite{KLS12, LP13}.  

Unfortunately, at this time, we are not prepared to offer an analytic treatment to explain the behavior of quantum walks in this topological context. Any attempt along these lines should prove both challenging and rewarding.

\end{document}